\documentstyle{article}
\begin{document}
\begin{center}
{\large{\bf Parametric up conversion of the vacuum }}\\
\vspace{1cm}
Trevor~W.~Marshall\\
{\small{\it Department of Mathematics, University of Manchester,
Manchester M13 9PL, U. K.}}\\
16, March, 1998
\end{center}
\begin{abstract}
The theory of parametric
down conversion of the vacuum,
based on a real zeropoint, or ``vacuum"
electromagnetic field, has been treated in
earlier articles. The same
theory predicts a hitherto unsuspected phenomenon --- parametric
up conversion
of the vacuum. This article describes how the phenomenon
may be demonstrated experimentally.\\
\noindent
PACS numbers: 03.65 42.50
\end{abstract}
\section{Introduction}
Parametric down conversion (PDC) and parametric up
conversion (PUC) have been known about since the
earliest days of nonlinear optics\cite{yariv,saleh}.
They occur when a light signal of frequency $\omega_1$
is incident on a nonlinear crystal which is pumped by a
laser of frequency $\omega_0$. In
PDC an idler of frequency $\omega_0-\omega_1$
is emitted; in PUC this emitted frequency is
$\omega_0+\omega_1$.

In the case of PDC the name
is rather anomalous, because the pumping converts
$\omega_1$ into $\omega_0-\omega_1$, which
can be either ``down" or ``up" relative to
$\omega_1$. The phenomenon has been given this
name because, in the most studied case,
the incident wave is a mode of the zeropoint or
``vacuum" field, and we then see {\it two coupled modes}
emerging from the crystal. Their frequencies are
$\omega_1$ and $\omega_2$ with $\omega_1+\omega_2=\omega_0$.
But the optical community has been reluctant to recognize
the reality of the zeropoint field; in the early
days of nonlinear optics it was sometimes called
``fictitious"\cite{kleinman}, but nowadays it is
not afforded even this status. So, according to the
current description, a ``photon" of the laser
with energy $\hbar\omega_0$ downconverts,
spontaneously, into a pair of ``photons" with
energies $\hbar\omega_1$ and $\hbar\omega_2$. If
we insist on the older, and in my view correct,
description, this latter phenomenon should,
properly, be called Parametric Down Conversion
of the Vacuum (PDCV). What is occurring is
the conversion of a zeropoint mode $\omega_1$
into a
detectable signal $\omega_2$, and
simultaneously the conversion of a zeropoint
mode $\omega_2$ into a signal $\omega_1$.

The zeropoint description of PDCV has been investigated
in a series of articles\cite{pdc1,pdc2,pdc3,pdc4}, where
we showed that it gives a consistent local
realist explanation for all PDCV phenomena
purporting to show photon entanglement. If we
change over from the photon description to the
wave description of light, all of these mysterious,
allegedly ``nonlocal" phenomena become local!

Now an obvious question to pose is ``If the
pump can convert a vacuum mode $\omega_1$ into a detectable
signal $\omega_0-\omega_1$ (PDCV), why is it that nobody
has reported seeing the conversion of $\omega_1$ into
$\omega_0+\omega_1$ (PUCV)?" My reply to this question
is simply that nobody has looked for this phenomenon;
if they look for it they will see it! All we need to
know is where to look and, at least approximately,
what intensity of signal to expect. We show how
to do this in the following sections.

\section{Position of the PUCV rainbow}
The direction of a PDCV (or PUCV) signal is
determined by the {\it phase matching condition}\cite{yariv},
which is that, at all points of the crystal and
at all times, the
phase of the pump coincides with the sum
(or difference) of the phases of the two coupled zeropoint
modes. In the case of PDCV this condition
may be translated into ``photon" language as
Conservation of Four-Momentum, but no such
translation exists for PUCV, which is possibly why
nobody has looked for PUCV. We shall suppose
the customary experimental setup, in which
the pump is normally incident on one of the
crystal faces. Then, if the refractive
indices of the modes ($\omega_0,\omega_1,\omega_2$)
are ($n_0,n_1,n_2$), and the wave vectors of
($\omega_1,\omega_2$) make angles ($\theta_1,\theta_2$)
with the pump, phase matching gives (upper signs are
PDCV, and lower are PUCV)
\begin{eqnarray}   
\omega_2&=&\omega_0\mp\omega_1\;,\\
\omega_2\sqrt{n_2^2-\sin^2\theta_2}&=&
\omega_0 n_0\mp\omega_1\sqrt{n_1^2-\sin^2\theta_1}\;,\\
\omega_2\sin\theta_2&=&\mp\omega_1\sin\theta_1\;.
\end{eqnarray}                                   
Typically, solution of these matching conditions, for $\theta_1$
and $\theta_2$ in terms of $\omega_1$, gives us PDCV and PUCV
as depicted in Figure 1.
\begin{figure}[htb]
\unitlength 0.35mm
\linethickness{0.4pt}
\begin{picture}(337.83,110)(0,70)
\put(70.00,140.34){\line(0,-1){40.00}}
\put(90.00,100.34){\line(0,1){40.00}}
\thicklines
\put(0.00,120.34){\line(1,0){70.00}}
\thinlines
\put(90.00,120.34){\line(5,1){69.33}}
\put(159.33,106.34){\line(-5,1){69.33}}
\put(17.33,115.34){\makebox(0,0)[cc]{laser}}
\put(80.00,120.34){\makebox(0,0)[cc]{{\small NLC}}}
\put(12.67,120.34){\vector(1,0){7.67}}
\put(14.67,131.34){\vector(4,-1){1.00}}
\put(119.33,114.34){\vector(4,-1){1.00}}
\put(119.67,126.01){\vector(4,1){1.00}}
\put(70.00,120.34){\line(-5,1){42.33}}
\put(15.67,131.34){\line(-5,1){15.67}}
\put(31.33,139.67){\makebox(0,0)[cc]{ZPF($\omega_1$)}}
\put(122.67,139.34){\makebox(0,0)[cc]{signal($\omega_2)$}}
\put(122.00,101.00){\makebox(0,0)[cc]{signal($\omega_1$)}}
\put(0.00,106.34){\vector(4,1){16.00}}
\put(70.00,120.34){\line(-5,-1){39.67}}
\put(31.33,101.67){\makebox(0,0)[cc]{ZPF($\omega_2$)}}
\put(248.50,162.67){\line(0,-1){80.00}}
\put(268.50,82.67){\line(0,1){80.00}}
\thicklines
\put(178.50,122.67){\line(1,0){70.00}}
\thinlines
\put(258.50,122.67){\makebox(0,0)[cc]{{\small NLC}}}
\put(191.17,122.67){\vector(1,0){7.67}}
\put(268.50,122.67){\line(5,-3){69.33}}
\put(337.83,81.07){\line(0,0){0.00}}
\put(210.83,116.34){\makebox(0,0)[cc]{laser}}
\put(298.50,105.00){\vector(3,-2){1.00}}
\put(300.17,117.34){\vector(4,-1){1.00}}
\put(206.17,148.00){\vector(3,-2){1.00}}
\put(211.83,162.34){\makebox(0,0)[cc]{ZPF($\omega_1)$}}
\put(248.50,122.67){\line(-6,1){30.67}}
\put(196.50,131.67){\line(-6,1){18.00}}
\put(191.50,132.34){\vector(4,-1){1.00}}
\put(203.50,138.00){\makebox(0,0)[cc]{ZPF($\omega_2$)}}
\put(309.83,127.67){\makebox(0,0)[cc]{signal($\omega_2$)}}
\put(294.17,90){\makebox(0,0)[cc]{signal($\omega_1$)}}
\put(248.50,122.67){\line(-5,3){27.00}}
\put(207.17,147.00){\line(-5,3){22.33}}
\put(268.50,122.67){\line(6,-1){62.33}}
\end{picture}
\caption{Typical PDCV (left) and PUCV(right) processes, as predicted by
classical electrodynamics incorporating a real zeropoint
field (ZPF). The latter modes have been denoted by interrupted
lines, indicating that they are below the intensity threshold for detection.
Note that the PDCV signal pair are on opposite sides of the pump,
while the PUCV pair are both on the same side.}
\end{figure}
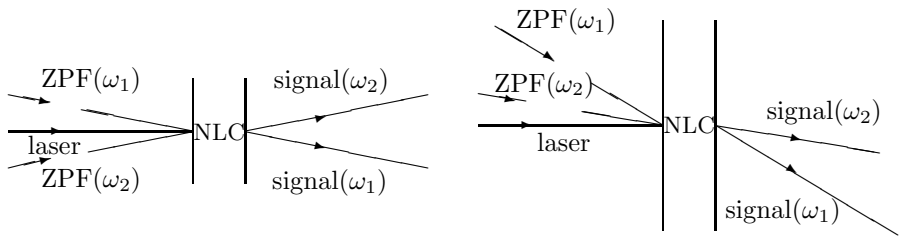

The wave vectors of the two outgoing modes
in Figure 1 depend on the refractive indices
of the crystal. We present here an experimental
design using the BBO crystal,
which is already widely used\cite{kwiat}
in PDC experiments, and
for which the ordinary and extraordinary
indices are\cite{ling} (wavelength in microns)
\begin{eqnarray}
n_{{\rm ord}}^2(\lambda)&=&2.7359+\frac{.01878}{\lambda^2-.01822}
-.01354\lambda^2\;,\\
n_{{\rm ext}}^2(\lambda,90)&=&2.3753+\frac{.01224}{\lambda^2-.01667}
-.01516\lambda^2\;,\\
\frac{1}{n_{{\rm ext}}^2(\lambda,\theta)}&=&
\frac{\cos^2\theta}{n_{{\rm ord}}^2(\lambda)}
+\frac{\sin^2\theta}{n_{{\rm ext}}^2(\lambda,90)}\;.
\end{eqnarray}

Suppose a BBO crystal is cut at $\theta=90$ degrees, which means
that its optic axis lies in one of its faces, and that
a pump of wavelength 351nm is normally incident on that face
and polarized in the ordinary direction, that is
perpendicular to the optic axis. Then from the
above matching conditions we may infer
that, for most of the visible spectrum,
that is for wavelengths $\lambda_1>481.07$nm, PUCV
occurs in the form of a rainbow,
that is each frequency
is emitted in a definite direction. It should be
emphasized that this direction is very different
from that of the PDCV rainbow asssociated with
$\omega_0$, and in any case the latter is
produced from a pump which is extraordinarily
polarized.

The frequency-angle dependence
of the lower-frequency mode is given in
the following Table. For a given frequency
the phase matching surface is a cone with its
axis in the direction of the pump. Its cross
section is an ellipse with the major axis
in the equatorial plane, and the minor
axis in the longitudinal plane, of the
index ellipsoid. I give the semiangle of
the cone in these two directions.
\begin{table}[htb]
\begin{center}
\begin{tabular}{llll}
wavelength(nm)&equatorial &
longitudinal&partner mode\\
&semiangle(deg)&semiangle(deg)&({\it e}-polarized)\\
481.07&0&0&202.93\\
500&18.04&15.37&206.23\\
600&42.42&36.94&221.45\\
700&55.98&49.18&233.78\\
800&68.13&59.47&243.96
\end{tabular}
\end{center}
\caption{
PUCV with a normally incident pump at $351o$. The
lower-frequency signal, which is $o$-polarized, is in
the three left-hand columns, while its
undetectable $e$-polarized partner is in the
right-hand column.
}
\end{table}

The title of {\it Up}-conversion refers in general
to the fact that the $e$-polarized partner has a
frequency higher than the pump, but, as we shall
see in the next section, this mode actually
has its intensity {\it reduced} below the
zeropoint level, so that it cannot be directly
detected.
In the above
example the whole of the $o$-polarized spectrum
associated with PUCV was below the pump
frequency, but I emphasize that, in the
zeropoint description, it is the ``vacuum"
rather than the pump which has its frequency
converted. That there is no reason why the
up-converted signal should necessarily
have a lower frequency than the pump may
be demonstrated by using a $702o$ instead of
a $351o$ pump, as shown in Table 2.
\begin{table}[htb]
\begin{center}
\begin{tabular}{llll}
wavelength(nm)&equatorial &
longitudinal&partner mode\\
&semiangle(deg)&semiangle(deg)&({\it e}-polarized)\\
256.79&0&0&188.01\\
270&28.04&16.99&195.00\\
300&45.57&28.69&210.18\\
400&65.08&44.23&254.81\\
500&73.16&51.69&292.01\\
600&79.67&56.93&323.50\\
679.5&89.33&60.47&345.28
\end{tabular}
\end{center}
\caption{
The same experimental setup as in Table 1, except that
the pump wavelength is 702nm.
}
\end{table}

The most convenient part of the PUCV spectrum is
probably those wavelengths which have
exiting angles in the range from about 10-30
degrees. I give, in Table 3, the wavelength
at the edge of the PUCV spectrum, where
the exiting angle is zero. I note an interesting
feature of this Table, namely that the partner mode
near the edge of the PUCV spectrum is, in all
cases, fairly close to, but above, the transparency limit of the
crystal (189nm).
\begin{table}[htb]
\begin{center}
\begin{tabular}{lll}
pump&edge of &
partner mode\\
wavelength&PUCV spectrum&({\it e}-polarized)\\
351&481.07&202.93\\
400&419.35&204.72\\
500&338.02&202.00\\
600&290.02&195.51\\
702&256.79&188.01
\end{tabular}
\end{center}
\caption{
The edge of the PUCV spectrum as a function of the
pump wavelength
}
\end{table}

The wavelengths used for an experimental
observation of PUCV will be close to, but rather greater than,
the edge wavelength. Obviously the most practical way to locate
this relatively weak PUCV signal is to use a second
laser suitably aligned. For example, with the
$351o$ pump, a $206.23e$ laser in the
equatorial plane, with an incidence
angle of 7.34 degrees, will result in a relatively
strong PUC signal at  500nm and 18.37 degrees.
Or, as an example where the up-converted mode
has a frequency higher than the pump, consider
a pump at 500nm. Then an aligning laser at
206.23nm, whose incidence angle is 12.71 degrees,
will produce a signal at 351nm exiting at
18.37 degrees.
The zeropoint theory predicts
that the intensity of the PUC signal
does not go to zero when this aligning laser is
removed.

\section{Intensity of the PUCV rainbow}
We now calculate the intensity of the PUCV
rainbow, using the crystal and pump geometry of
the previous section. It is simplest to consider
the signal emitted in the equatorial plane of
the index ellipsoid, which we denote as the
$xy$-plane; the pump's wave vector is
($\omega_0,0,0$) and its polarization direction is
$(0,1,0)$. So the signal has wave  vector
$(\omega_1\cos\theta_1,\omega_1\sin\theta_1,0)$
and polarization direction $(-\sin\theta_1,\cos\theta_1,0)$,
while its partner mode has wave vector
($\omega_2\cos\theta_2,\omega_2\sin\theta_2,0$) and
polarization direction $(0,0,1)$. We denote the magnitudes of the
electric vectors of the three coupled modes as
($E_0,E_1,E_2$). Then the relevant crystal
polarizations are
\begin{eqnarray}
P_1&=&\frac{n_1^2-1}{4\pi}E_1+2d_{15}E_0E_2\cos\phi_1\:,\\
P_2&=&\frac{n_2^2-1}{4\pi}E_2+2d_{31}E_0E_1\cos\phi_1\;,
\end{eqnarray}
where $\phi_i$ gives the direction of a signal
inside the crystal, that is
\begin{equation}
\sin\phi_i=\frac{\sin\theta_i}{n_i}\;,\label{thetau}
\end{equation}
and $d_{15},d_{31}$ are the appropriate
second-order polarizabilities.
Note that, with our choice
of geometry, we have simply
\begin{equation}
n_0=n_{{\rm ord}}(\omega_0)\;,\;
n_1=n_{{\rm ord}}(\omega_1)\;,\;
n_2=n_{{\rm ext}}(\omega_2,90)\;.
\end{equation}
We now make a linearizing approximation which is
equivalent to
neglecting the depletion of the pump inside the crystal,
namely we put
\begin{equation}
E_0=V\cos(\omega_0t-k_0x)\;,\;(k_0=n_0\omega_0)\;,
\end{equation}
and treat $V$ as constant.
We also neglect dissipation within the crystal,
and define 
\begin{equation}               
f_1=4\pi d_{15}V\cos\phi_1\;,\;f_2=4\pi d_{31}V\cos\phi_1\;.
\end{equation}
Then the Maxwell equations
coupling $E_1$ and $E_2$ are
\begin{eqnarray}
(\Delta+n_1^2\partial^2/\partial t^2)E_1&=&-f_1
(\partial^2/\partial t^2)E_2
\cos(i\omega_0t-ik_0x)\;,\\
(\Delta+n_2^2\partial^2/\partial t^2)E_2&=&-f_2
(\partial^2/\partial t^2)
E_1\cos(i\omega_0t-ik_0x)\;.
\end{eqnarray}
Now we substitute
\begin{equation}
E_i(x,y,z)=\sqrt{\frac{\omega_i}{n_i}}[A_i(x)e^{i\omega_i't
-ik_ix-\omega_i\sin\theta_i y}+{\rm c.c.}]\;,
\end{equation}
where
\begin{equation}
\omega_1'\approx\omega_1\quad,\quad \omega_2'=\omega_0+\omega_1'
\quad {\rm and}\quad k_i=\sqrt{n_i^2\omega_i'^2
-\omega_i^2\sin^2\theta_i}\;,\label{spectrum}
\end{equation}
and put
\begin{equation}
g_i=\frac{f_i}{2\cos\phi_i}\sqrt{\frac{\omega_1\omega_2}{n_1n_2}}\;,
\label{gdef}
\end{equation}
and we make the slowly-varying-envelope approximation,
that is we discard second derivatives of $A_i$.
Then the coupling equations become
\begin{eqnarray}
\frac{dA_1}{dx}&=&-ig_1A_2e^{i\Delta x}\;,\\
\frac{dA_2}{dx}&=&-ig_2A_1e^{-i\Delta x}\;,
\end{eqnarray}
where
\begin{equation}
\Delta=k_0+k_1-k_2\;.
\end{equation}
Note that, when $\omega_1'=\omega_1$, we have perfect
phase matching, and then
$\Delta=0$. These
equations are the generalization of Ref.\cite{yariv},
eq.(8.7-2); $g_1$ and $g_2$ differ because we are not
putting $f_1=f_2$, and also because the rays are not
all collinear.

For a crystal of length $l$ we may solve
the above coupling equations and obtain the
mode amplitudes at $x=l$ in terms of those
at $x=0$:
\begin{eqnarray}
A_1(l)e^{-i\Delta x/2}&=&A_1(0)[\cos(bl)-i(\Delta l/2){\rm sinc}(bl)]
-ig_1lA_2(0){\rm sinc}(bl)\;,\\
A_2(l)e^{i\Delta x/2}&=&A_2(0)[\cos(bl)-i(\Delta l/2){\rm sinc}(bl)]
-ig_2lA_1(0){\rm sinc}(bl)\;,
\end{eqnarray}
where
\begin{equation}
{\rm sinc}(x)=\frac{\sin x}{x}\quad {\rm and} \quad
b=\sqrt{(\Delta^2/4)+g_1g_2}\;.
\end{equation}
The corresponding relation between the $x$-components of
the Poynting vectors, that is ${\cal P}_i=k_iA_iA_i^*/n_i$, is
\begin{eqnarray}
{\cal P}_1(l)=[\cos^2(bl)+(\Delta^2l^2/4){\rm sinc}^2(bl)]
{\cal P}_1(0)+(k_1n_2g_1^2l^2/k_2n_1){\rm sinc}^2(bl){\cal P}_2(0)\;,\\
{\cal P}_2(l)=[\cos^2(bl)+(\Delta^2l^2/4){\rm sinc}^2(bl)]
{\cal P}_2(0)+(k_2n_1g_2^2l^2/k_1n_2){\rm sinc}^2(bl){\cal P}_1(0)\;.
\end{eqnarray}
In order to obtain the relation between the
Poynting vectors of the incoming and outgoing
waves, we need, in addition to the latter
relations, the transmission coefficients
at the crystal interfaces, which are\cite{jackson}
\begin{equation}
\frac{{\cal P}_i(0)}{{\cal P}_i({\rm in})}=
\frac{{\cal P}_i({\rm out})}{{\cal P}_i(l)}=1-r_i\;,
\end{equation}
where
\begin{equation}
r_1=\frac{\tan^2(\theta_1-\phi_1)}{\tan^2(\theta_1
+\phi_1)}\;,\;
r_2=\frac{\sin^2(\theta_2-\phi_2)}{\sin^2(\theta_2
+\phi_2)}\;.
\end{equation}
Combining these two sets of relations, we obtain
\begin{eqnarray}
{\cal P}_1^{(0)}({\rm out})=[1-g_1g_2l^2{\rm sinc}^2(bl)]
(1-r_1)^2
(\hbar k_1/2n_1)\qquad\qquad\nonumber\\
+g_1^2l^2{\rm sinc}^2(bl)
(1-r_1)(1-r_2)
(\hbar k_1/2n_1)\;,\\
{\cal P}_2^{(0)}({\rm out})=[1-g_1g_2l^2{\rm sinc}^2(bl)]
(1-r_2)^2
(\hbar k_2/2n_2)\qquad\qquad\nonumber\\
+g_2^2l^2{\rm sinc}^2(bl)
(1-r_1)(1-r_2)
(\hbar k_2/2n_2)\;,
\end{eqnarray}
where we have used the standard
zeropoint spectrum, for which the Poynting
vector of a mode {\bf k}
in free space is $(\hbar{\bf k}/2)$.
The superscript (0) indicates that we have
calculated the outgoing Poynting vectors
after zero  reflections.
The contributions from a single reflection all
originate in zeropoint modes coming,
in Fig.2, from the right side of the crystal,
and they give, to order $g_{1,2}^2$,
\begin{eqnarray}
(2n_1/\hbar k_1){\cal P}_1^{(1)}({\rm out})=
r_1+r_1(1-r_1)^2
[1-g_1g_2l^2{\rm sinc}^2(bl)]
\nonumber\\
+r_2(1-r_1)(1-r_2)g_1^2l^2{\rm sinc}^2(bl)
\;.
\end{eqnarray}
with a similar equation for ${\cal P}_2^{(1)}({\rm out})$.
So zero and one reflections together give
\begin{eqnarray}
(2n_1/\hbar k_1){\cal P}_1^{(0+1)}({\rm out})=
(1-r_1^2+r_1^3)
[1-g_1g_2l^2{\rm sinc}^2(bl)]
\nonumber\\
+[(1-r_1-r_2^2+r_1r_2^2)g_1^2+r_1g_1g_2]{\rm sinc}^2(bl)
\;.        
\end{eqnarray}
The contributions from two reflections originate,
as do those with zero reflections, in modes coming
from the left. They should, strictly speaking, be
superposed, but that would suppose what is probably
an impossible accuracy in cutting the crystal, so
we shall simply add their Poynting vector also,
that is
\begin{eqnarray}
(2n_1/\hbar k_1){\cal P}_1^{(2)}({\rm out})=
r_1^2(1-r_1)^2
[1-g_1g_2l^2{\rm sinc}^2(bl)]
\nonumber\\
+(r_1^2+r_2^2)(1-r_1)(1-r_2)g_1^2{\rm sinc}^2(bl)
\;,
\end{eqnarray}
giving
\begin{eqnarray}
(2n_1/\hbar k_1){\cal P}_1^{(0+1+2)}({\rm out})=
(1-r_1^3+r_1^4)
[1-g_1g_2l^2{\rm sinc}^2(bl)]
\nonumber\\
+[\{1-r_1+r_1^2-r_1^3-r_2(1-r_1)(r_1^2+r_2^2)\}g_1^2+(r_1-r_1^2)g_1g_2]{\rm sinc}^2(bl)
\;.
\end{eqnarray}
The extension to all reflections is a straightforward
piece of algebra\cite{puc1}. We give the result
\begin{eqnarray}
{\cal P}_1^{(0+1+2+\ldots)}({\rm out})=
\frac{\hbar k_1}{2n_1}\left[1+\frac{g_1(g_1-g_2)l^2}
{1+r_1}{\rm sinc}^2(bl)\right]
\;,\nonumber\\
{\cal P}_2^{(0+1+2+\ldots)}({\rm out})=
\frac{\hbar k_2}{2n_2}\left[1+\frac{g_2(g_2-g_1)l^2}
{1+r_2}{\rm sinc}^2(bl)\right]
\;.
\end{eqnarray}

The detection theory appropriate for a semiclassical
theory like ours, which incorporates a real zeropoint
field, is different from most semiclassical theories
previously considered\cite{mandel}, because we
suppose that detectors are activated only by signals
whose intensity is above the zeropoint intensity. For
details we refer to Ref.\cite{pdc4}. So the photocount
rate ${\cal D}_i$, in the case of PUCV, will depend on the sign of
$g_1-g_2$. For the case $g_1>g_2$ we will have
\begin{equation}
{\cal D}_2=0\;,\;{\cal D}_1=
{\cal C}(\omega_1)\frac{\hbar k_1\omega_1^2}{2n_1}
\int_{\Delta\Omega_1}\int_{\Delta\omega_1}
\frac{g_1(g_1-g_2)l^2}
{1+r_1}{\rm sinc}^2(bl)d\omega_1'd\Omega_1\;,
\label{detpuc}
\end{equation}
and, for $g_1<g_2$,
\begin{equation}
{\cal D}_1=0\;,\;{\cal D}_2=
{\cal C}(\omega_2)\frac{\hbar k_2\omega_2^2}{2n_2}
\int_{\Delta\Omega_2}\int_{\Delta\omega_2}
\frac{g_2(g_2-g_1)l^2}
{1+r_2}{\rm sinc}^2(bl)d\omega_2'd\Omega_2\;,
\end{equation}
where ${\cal C}(\omega_i)$ is a detector efficiency,
$\Delta\Omega_i$ is the solid angle subtended by the
detector at the crystal, and $\Delta\omega_i$ is the
bandwidth which the detector accepts.

We have shown, therefore, that of the two ``entangled"
partners $\omega_1$ and $\omega_2$, only one is
directly detectable; its partner is depleted
below the zeropoint level and can not be detected.
It seems plausible to suppose that the depleted
partner is always $\omega_2$, which is
what happens when that mode is the pump. For this
to be the case at the edge of the PUCV spectrum,
displayed in Table 3 of the previous section,
would require that $d_{15}\geq d_{31}$. We shall
return to this question in a moment.

In the case of PDCV the whole calculation follows
the same lines, and the result is the same
except for certain changes of sign. We obtain
\begin{eqnarray}
{\cal D}_{1D}&=&{\cal C}(\omega_{1D})
\;\frac{\hbar k_{1D}\omega_{1D}^2}{2n_{1D}}
\int_{\Delta\Omega_{1D}}\int_{\Delta\omega_{1D}}
\qquad\qquad\qquad\qquad\qquad\nonumber\\
&&\frac{g_{1D}(g_{1D}+g_{2D})l^2}
{1+r_{1D}}{\rm sinc}^2({b_D}l)d\omega_{1D}'d\Omega_{1D}\;,\\
{\cal D}_{2D}&=&{\cal C}(\omega_{2D})
\;\frac{\hbar k_{2D}\omega_{2D}^2}{2n_{2D}}
\int_{\Delta\Omega_{2D}}\int_{\Delta\omega_{2D}}
\qquad\qquad\qquad\qquad\qquad\nonumber\\
&&\frac{g_{2D}(g_{1D}+g_{2D})l^2}{1+r_{2D}}
{\rm sinc}^2(b_Dl)d\omega_{2D}'d\Omega_{2D}\;,
\end{eqnarray}
where
\begin{equation}
b_D=\sqrt{(\Delta^2_D/4)-g_{1D}g_{2D}}\quad{\rm and}\quad
\Delta_D=k_0-k_1-k_2\;.
\end{equation}
These latter variables, in the PUCV case, are
\begin{equation}
b=\sqrt{(\Delta^2/4)+g_1g_2}\quad{\rm and}\quad
\Delta=k_0+k_1-k_2\;.
\end{equation}
Since, in both cases, $g_1g_2l^2<<1$, we may put
simply
\begin{equation}
b=\Delta/2\quad{\rm and}\quad b_D=\Delta_D/2\;.
\end{equation}

Two detection procedures are commonly used; in one
of them $\Delta\Omega$ is
narrowly determined
by a ``pinhole" iris, while all frequencies are
accepted by the detector; in the other the detector
accepts only radiation which has passed through
a narrow frequency filter $\Delta\omega$, but collects over a
wide, effectively infinite, angular range. These
procedures are completely equivalent, but the
analysis used above is more easily continued
in terms of the first one. An integration
of the $\Delta$-dependent part of the intensity,
using the frequency dependence of $\Delta$
given by eq.(\ref{spectrum})
gives, approximately,
\begin{equation}
\int_0^\infty{\rm sinc}^2(\Delta l/2)d\omega_1'=
\frac{2\pi}{l\mid n_1\sec\phi_1-n_2\sec\phi_2\mid}\;,
\label{freqint}
\end{equation}
and in the PDCV case we obtain
\begin{equation}
\int_0^\infty{\rm sinc}^2(\Delta_Dl/2)d\omega_1'=
\frac{2\pi}{l\mid n_{1D}\sec\phi_{1D}-
n_{2D}\sec\phi_{2D}\mid}\;.
\end{equation}
Both of the latter approximations are good,
provided we do not go too near the maximum,
which occurs when the denominator on the
right hand side is zero. There a more
sophisticated approximation is required\cite{kleinman};
the integral is proportional to $l^{-1/2}$ instead
of $l^{-1}$ and dissipation can no longer be neglected.

I now want to compare the counting rate in a PUCV
experiment, in which  a pinhole is placed at an
angle $\theta_1$ corresponding to a signal
whose frequency is near $\omega_1$, with a PDCV
experiment designed to detect a signal near
$\omega_{1D}$, whose pinhole is at $\theta_{1D}$.
Let us assume that both irises
collect over the same solid angle.
Then, substituting eqs.(\ref{gdef},\ref{freqint}) in eq.(\ref{detpuc}),
the counting rate for PUCV is
\[
{\cal D}_1=16\pi^3\hbar V^2l{\cal C}(\omega_1)
\Delta\Omega\;\frac{\omega_1^4(\omega_0+\omega_1)}{n_1n_2}
\;\frac{d_{15}(d_{15}\sec\phi_1
-d_{31}\sec\phi_2)}{\mid n_1\sec\phi_1-n_2\sec\phi_2\mid}
\;\cdot\qquad\qquad
\]
\begin{equation}
\qquad\qquad\qquad\cdot\;
\frac{\cos^2\phi_1\tan^2(\theta_1+\phi_1)}
{\tan^2(\theta_1+\phi_1)+\tan^2(\theta_1-\phi_1)}\;.
\label{pucrate}
\end{equation}
The corresponding rate for ($e\rightarrow o+o$, equatorial) PDCV is
\[
{\cal D}_{1D}=16\pi^3\hbar V^2l{\cal C}(\omega_{1D})
\Delta\Omega\;\frac{\omega_{1D}^4(\omega_0-\omega_{1D})}{n_{1D}n_{2D}}
\;\frac{d_{15D}^2(\sec\phi_{1D}+\sec\phi_{2D})}
{\mid n_{1D}\sec\phi_{1D}-n_{2D}\sec\phi_{2D}\mid}
\;\cdot\qquad\qquad
\]
\begin{equation}
\qquad\qquad\cdot\;
\frac{\cos^2(\phi_{1D}-\phi_{2D})\tan^2(\theta_{1D}+\phi_{1D})}
{\tan^2(\theta_{1D}+\phi_{1D})+\tan^2(\theta_{1D}-\phi_{1D})}\;.
\end{equation}

Note that we should distinguish between $d_{ij}$ and $d_{ijD}$,
because these coupling constants depend on the participating
mode frequencies. However, we do not yet have detailed
information about such frequency dependence, so I shall
now make the rather gross approximation that
\begin{equation}
d_{ijk}\;{\rm is\;independent\;of\;frequency\;and\;}d_{ijk}=d_{jik}\;,
\label{kleinconj}
\end{equation}
so that $d_{15},d_{31}$ and $d_{15D}$ are all equal.
The second part of this approximation,
known generally as the {\it Kleinman conjecture}\cite{yariv1},
may be shown to be a consequence of the first.
It is sometimes stated as
a ``theorem" (Ref.\cite{saleh}, page 780),
despite some experimental evidence (Ref.\cite{yariv}, Table 8.1
and Ref.\cite{saleh}, Table 19.6-1) that $d_{ijk}$ and $d_{jik}$
can differ by up to 20 per cent.
All that can safely be said is that
the {\it approximate} frequency dependence of $d_{ijk}$
incurs the {\it approximate} validity of the Kleinmann
conjecture, which justifies its description\cite{yariv1} as
``a powerful practical relationship". Note that this
approximation gives $g_1=g_2$ in the case of collinear PUCV,
so that the intensity is then zero at the edge of the PUCV
spectrum. Also, in the same spirit, we shall put
${\cal C}(\omega)=$ constant. Note that the quantity
called ``quantum efficiency" is $\hbar\omega{\cal C}(\omega)$.

The PDCV process which we take as a
basis of comparison uses a $351e$
pump down-converting into two ordinarily
polarized partners, one of which is $692o$
which corresponds to $\omega_{1D}$.
That brings us fairly close to the maximum
counting rate for PDCV, which occurs for the
symmetrical case $\lambda_1=702$nm.
The ratio of ${\cal D}(\omega_1)$ to ${\cal D}(\omega_{1D})$
has been calculated, for various $\omega_1$,
using the approximations
indicated in the previous paragraph,
and the results are displayed in Table 4.
\begin{table}[htb]
\begin{center}
\begin{tabular}{llll}
wavelength(nm)&${\cal D}(\omega_1)/{\cal D}(\omega_{1D})$&
$\theta_1$(degrees)\\
482&0.003&4.07\\
484&0.011&7.20\\
486&0.025&9.33\\
488&0.059&11.04\\
490&0.254&12.50\\
492&0.221&13.81\\
494&0.094&14.99\\
496&0.065&16.08\\
498&0.052&17.09\\
500&0.045&18.04
\end{tabular}
\end{center}
\caption{
Ratio of PUCV to a standard PDCV counting rate as a function
of wavelength.
}
\end{table}

We conclude that there is a peak
in the intensity of the PUCV rainbow near
491nm, with a maximum counting rate around
30 per cent of the standard PDCV process.
Of course, this PUCV maximum is
nothing like as strong as the PDCV maximum at
702nm, but the above calculation indicates that
it will nevertheless be easy to observe.
Furthermore, the size of the maximum is
extremely sensitive to a small departure
from the approximation (\ref{kleinconj});
if $d_{31}$ is 5 per cent less than $d_{15}$
it increases to somewhere between 100 and
150 per cent of the standard PDCV value.
\section{Conclusion}
We have shown, in the previous two sections, that PUCV,
according to the theory of a real zeropoint field, must
occur, and that it should
not be too difficult to detect. Of course, observation of
this phenomenon will be strong evidence in favour of a real
ZPF.

We showed, in our earlier series of articles\cite{pdc1,pdc2,pdc3,pdc4},
that the extremely problematic and nonlocal concept
of {\it photon entanglement} between a coupled
pair of PDCV signals receives a local and fully
causal explanation in terms of a real ZPF. In the
ZPF description, entangled photons are waves whose
amplitudes are correlated below as well as above
the level at which all detector thresholds are
set. Now we find that, in PUCV, there is still
a pair of entangled signals, but that one of them,
that with its frequency above the pump frequency,
is entirely below threshold, and therefore cannot be
detected. Once PUCV has been experimentally
demonstrated it will be a task for nonlinear
opticians to devise ways and means of
demonstrating this subzeropoint entanglement.
One possible way to reveal the ghostly
high-frequency partner would be to look
for its mirror image emerging from the
crystal in the backward direction. According
to the analysis of the previous section
this mode has an intensity above threshold,
but the counting rate is reduced by a
factor of $r_2$ relative to its
forward low-frequency partner.

\end{document}